# Rolling bearing fault diagnosis method based on generative adversarial enhanced multi-scale convolutional neural network model

Maoxuan Zhou[1], Wei Kang[2] and Kun He[2]

[1] Kunming University of Science and Technology, Faculty of Mechanical and Electrical Engineering, China

[2] Corresponding author email: kang_wei
@stu.xjtu.edu.cn

**Abstract.** In order to solve the problem that current convolutional neural networks can not capture the correlation features between the time domain signals of rolling bearings effectively, and the model accuracy is limited by the number and quality of samples, a rolling bearing fault diagnosis method based on generative adversarial enhanced multi-scale convolutional neural network model is proposed. Firstly, Gram angular field coding technique is used to encode the time domain signal of the rolling bearing and generate the feature map to retain the complete information of the vibration signal. Then, the resulting data is divided into a training set, a validation set, and a test set. Among them, the training set is input into the gradient penalty Wasserstein distance generation adversarial network to complete the training, and a new sample with similar features to the training sample is obtained, and then the original training set is expanded. Next, multi-scale convolution is used to extract the fault features of the extended training set, and the feature graph is normalized by example to overcome the influence of the difference in feature distribution. Finally, the attention mechanism is applied to the adaptive weighting of normalized features and the extraction of deep features, and the fault diagnosis is completed by the softmax classifier. Compared with ResNet method, the experimental results show that the proposed method has better generalization performance and anti-noise performance.

## 1 Introduction

The running state of rolling bearings is closely related to the working state of mechanical equipment[1]. Rolling bearing failure, light will cause economic losses, heavy may cause casualties[2]. Therefore, it is crucial to accurately monitor the operating status of rolling bearing[3].

At present, the fault analysis of rolling bearing mainly includes signal acquisition, data preprocessing, feature extraction and classification recognition. Traditional fault diagnosis methods mainly rely on empirical judgment, and the process of feature extraction is not only time-consuming, but also the obtained feature information is not comprehensive enough[4]. With the rapid development of computer technology, machine learning algorithms such as support vector machine and BP neural network are introduced into the field of fault diagnosis[5, 6]. It should be noted that these algorithms belong to the shallow network domain, and the feature information they can extract is limited, so their fault diagnosis accuracy is relatively low. To overcome these problems, deep learning algorithms have gained widespread attention in recent years. Deep learning algorithm can automatically extract features through multi-layer neural network, which can capture higher-level and richer feature information. This method can realize more accurate fault diagnosis by learning the characteristics of a large number of data samples. For example, a convolutional neural network[7] (CNN) can efficiently process image data, a recurrent neural network[8] (RNN) can process time series data, and a converter model[9] (Transformer) can process sequence data.

The above method has the following problems in engineering application: 1. It is difficult and costly to obtain actual fault signals, and it is difficult to achieve high-precision fault identification through small-sample training. 2. Environmental noise has a great impact on recognition accuracy. The methods of data expansion and transfer learning are usually adopted to solve the small sample problem.

One solution to this problem is data augmentation. GOODFELLOW[10] et al. proposed the generative adversarial network (GAN), which has a remarkable ability in image generation. The application of generative adversarial network is remarkable in view of the small number of equipment fault samples and data imbalance. wang[11] et al. proposed a fault diagnosis method for GAN and stacked autoencoder. ding[12] et al. proposed a small sample mechanical fault diagnosis method based on GAN. Although the above methods have a certain contribution to fault sample expansion, they do not give full play to the image generation capability of GAN. Xiao Xiong[13] et al. proposed a method to transform one-dimensional time signal into two-dimensional feature image and realize data enhancement with the help of generative adversarial network. However, this method ignores the correlation of signals in time.

The Gram angular difference field (GADF) presents an innovative representation of time series. By mapping the single dimension signal to the extreme coordinate system and performing the dot product operation, it realizes the Angle differentiation between different modes at the same time point, thus maintaining the difference between modes and ensuring the maximum distinguisability between features. Next, the one-dimensional time series is encoded by the Gram matrix (GM) generated by the dot product operation. In the GM matrix, the elements gradually increase with time, their values follow a Gaussian distribution, and the time information is effectively encoded in the resulting image. In addition, the obtained two-dimensional images show high sparsity, effectively eliminating redundant information. Therefore, GADF images not only contain the numerical information of one-dimensional time series, but also retain the relationship of time information in non-stationary time-varying signals, the difference between multiple modes, and the correlation of potential states, and reduce the redundant information. GADF technology can be used to maximize the distinction between features, which provides an effective means for deep feature description of broadband oscillations.

The coupling components in the vibration signal are usually different, which contain multiple vibration modes, have multi-scale characteristics, and are more complex on the time scale. Traditional CNN reduces the dependence on prior knowledge, but there is only one convolution kernel, so it is difficult to extract multi-scale features [14]. For traditional CNNS, multi-scale CNNS have a wider field of view and generalization ability, and can analyze and extract global information and local features of data at the same time. Zhang Mingde[15] et al. proposed a multi-scale convolution method and verified its feasibility with open data. Peng Peng[16] et al. proposed a fault diagnosis method of rotary reducer under the influence of noise using multi-scale convolution. Shen Changqing[17] et al. proposed an adaptive transfer learning model based on multi-scale convolution, which effectively solved the bearing fault mode under varying working conditions.

Based on the above problems, this paper combines Gramian Angular Fields (GAF) and gradient penalty Wasserstein Range Generation Adversary-Network (WGAN-GP). The Gramian Angular Difference Field (GADF) is used to convert the time domain signal into a two-dimensional feature map, which can better preserve the integrity of the time signal. The obtained feature map is input into WGAN-GP to give full play to its image generation capability. Finally, the obtained data are input into Multiscale attention convolutional neural networks (MACNN) to achieve fault identification.Theoretical part

## 2 Theoretical Part

### 2.1 Gram matrix

Gram matrix (GM) builds the inner product between self-vectors, which is equivalent to the concentration covariance matrix of different data features, and its calculation formula is as follows:

$$\triangle(\alpha_1, \alpha_2 \cdots \alpha_n) = \begin{pmatrix} (\alpha_1, \alpha_1) & (\alpha_1, \alpha_2) & \cdots & (\alpha_1, \alpha_n) \\ (\alpha_2, \alpha_1) & (\alpha_2, \alpha_2) & \cdots & (\alpha_2, \alpha_n) \\ \vdots & \vdots & \vdots & \vdots \\ (\alpha_n, \alpha_1) & (\alpha_n, \alpha_2) & \cdots & (\alpha_n, \alpha_n) \end{pmatrix} \quad (1)$$

In the above formula, $\alpha_i \{i = 0, 1, \cdots n\}$ represents different eigenvectors, $()$ is the inner product operation.

In the matrix n*n obtained by the inner product, different elements are inner products of different vectors. As shown in Figure 1 below:

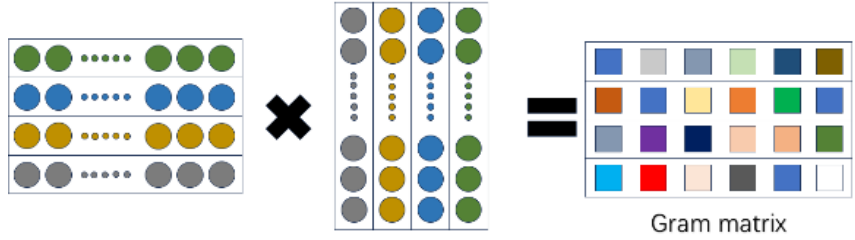

Fig.1 GM diagram

Therefore, GM not only shows the number of features, but also reflects the close relationship between the different features. The GM calculated by the feature vector can reveal the potential relationship between the image features. If the GM difference between two images is small, then the two images are considered similar.

### 2.2 GADF

Compared with RP and MTF image coding techniques, the Gram Angle difference field transforms the correlation of each one-dimensional signal point in different time intervals by Angle difference[18], visualizes through pixel points, and enhances the fault characteristics of one-dimensional signal[19]. Gram angular difference field realization steps:

①Time series signals for each set of samples $T = \{t_1, t_2, ..., t_n\}$ use normalization scale to $[-1, 1]$, as shown in equation (2):

$$\tilde{t} = \frac{(t_i - \min(T)) + (t_i - \min(T))}{\max(T) - \min(T)} \quad (2)$$

②Convert the time series to polar coordinates:

$$\begin{cases} \varphi = \arccos(\tilde{t}_i), -1 \leq \tilde{t}_i \leq 1, \tilde{t}_i \in \tilde{T} \\ a_i = \frac{x_i}{M}, x_i \in M \end{cases} \quad (3)$$

In the above formula: $\varphi$ represents the Angle corresponding to the cosine polar coordinates, $a$ represents the time period, and $M$ is a constant factor. Formula (3) divides the interval of length 1 into parts of quantity M and equal size, $a_i$ indicating that the radius is accumulated to $i$.

③The formula for the Gram difference field is as follows:

$$G_{GADF} = \begin{bmatrix} \sin(\varphi_1 - \varphi_1) & \sin(\varphi_1 - \varphi_2) & \cdots & \sin(\varphi_1 - \varphi_n) \\ \sin(\varphi_2 - \varphi_1) & \sin(\varphi_2 - \varphi_2) & \cdots & \sin(\varphi_2 - \varphi_n) \\ \vdots & \vdots & \vdots & \vdots \\ \sin(\varphi_n - \varphi_1) & \sin(\varphi_n - \varphi_2) & \cdots & \sin(\varphi_n - \varphi_n) \end{bmatrix}$$

$$= \left[ \sin(\varphi_i - \varphi_j) \right] = \sqrt{A - \tilde{B}^2} * B - \tilde{B} * \sqrt{A - B^2} \quad (4)$$

In the above formula, $A$ is the unit row vector, $B$ and $\tilde{B}$ is the row vector before and after scaling, and $\varphi_i$ is the Angle between the vectors.

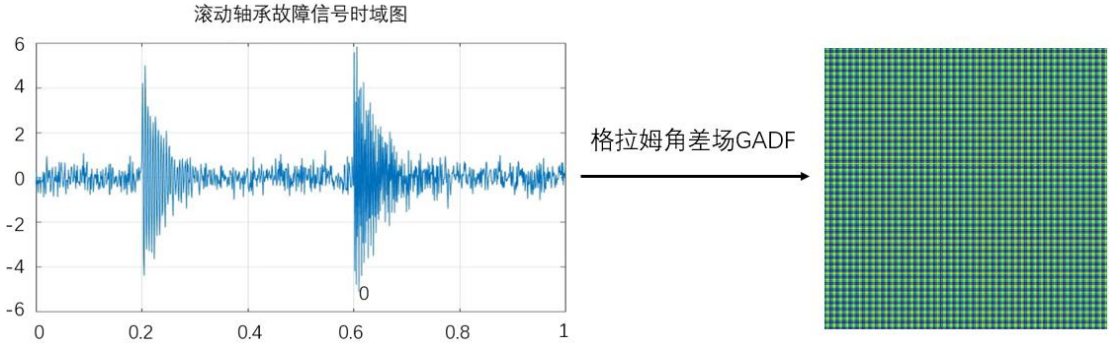

Fig.2 Gram image coding technique

## 2.3 WGAN-GP

First proposed in game theory, Gans are dual network models consisting of discriminators and generators. First, the noise is input into the generator module, and the generator G (z) generates a sample similar to the real sample by learning from the real sample. Then the discriminator D (z) outputs a probability P(z) by comparing the real sample with the generated sample, which represents the similarity probability of the real sample and the generated sample. In the training process, the generator learns the feature distribution of the real sample, so that the discriminator can improve the ability to distinguish the generated sample from the real sample. The formula for traditional GAN is as follows:

$$\max_D \min_G D(G, D) = E_{Z \sim G_{Z(Z)}} \left[ \lg(1 - D(G(Z))) \right] + E_{y \sim P_{a(y)}} \left[ \lg(D(y)) \right] \quad (5)$$

The above formula, $G_{Z(Z)}$ and $P_{a(y)}$, in turn, are Gaussian distributions and true sample distributions.

In the course of training, the probability $P(z)$ of the discriminator gradually approaches 1 and $D(G(Z))$ gradually approaches 0. The probability $D(G(Z))$ of the generator gradually approaches 1.

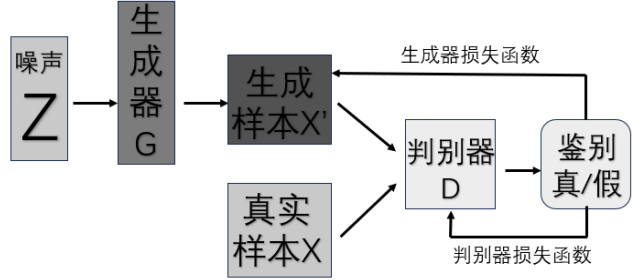

Fig.3 GAN basic model

When the distribution of traditional GAN is not overlapping, the gradient will disappear, resulting in the situation of local optimal solution. To solve this problem, Wasserstein distance[20] is added to traditional GAN. The formula is as follows:

$$L(P_i, P_j) = \text{INF}_{a \sim \prod(P_i, P_j)} E_{(m,n) \sim a} \left[ \|m - n\| \right] \qquad (6)$$

Its structure is restricted by Lipschitz condition, and due to the constraint of continuity, its continuous function changes as follows:

$$f(m) - f(n) \leq K|m - n| \qquad (7)$$

GAN based on Wasserstein distance is shown as follows:

$$l(G, D)_{|D| \leq k} = E_{a \sim pg(z)}[D(z)] - E_{x \sim pd(x)}[D(x)] \qquad (8)$$

According to the above formula, the weight reduction problem will cause the gradient to explode or disappear. Therefore, introducing gradient penalty[21] into WGAN can make model training more stable. WGAN-GP is shown as follows:

$$l(G, D) = E_{a \sim pg(z)}[D(z)] - E_{x \sim pd(x)}[D(x)] + \chi E_{x \sim px}(\|\nabla D(\hat{x})\|_2 - 1)^2 \qquad (9)$$

In the above formula, $\hat{x}$ is the random sample penalty factor.

**2.4  Instance normalization**

Instance normalization[22] (IN) was first proposed by ULYANOV on CVPR, which is used to normalize feature images to reduce differences and make up for the deficiency of BN. Its mathematical expression is as follows:

$$\begin{cases} \bar{M}^{L(C)} = \dfrac{M^{L(C)} - \eta_c}{\sqrt{\lambda_C^2 + \nu}} \\ Z^{L(C)} = \gamma^{L(C)} \bar{M}^{L(C)} + \alpha^{L(C)} \end{cases} \qquad (10)$$

In the above formula, $M^{L(C)}$ is the characteristic diagram of the L-th channel; $\bar{M}^{L(C)}$ is the normalized image, $v$ is the constant factor, $\gamma^{L(C)}$ and $\alpha^{L(C)}$ is the scale coefficient and the migration constant.

## 2.5 Convolutional neural network

CNN, as a standard feedforward neural network in the field of deep learning, converts simple features into more complex features through convolution operations, adopts hierarchical learning strategies, and has excellent feature representation capabilities. Its advantages lie in reducing the number of parameters required for training and mitigating the risk of network overfitting. Generally, the CNN structure includes convolution layer, pooling layer, activation layer and fully connected layer, etc., which can extract features efficiently for classification tasks.

Convolution layer extracts local features by applying convolution kernel to local input region, and the number of convolution kernel in the layer determines the number of channels to generate feature map. Its mathematical expression is as follows:

$$y_k^j = \sum_i^n x_i^{j-1} * w_k^j + b_k^j \qquad (11)$$

In the above formula, $x_i^{j-1}$ indicates input at layer $j$, $w_k^j$ is the weight of the convolution kernel, $b_k^j$ is the convolution kernel bias, $y_k^j$ is output at layer $j$.

The pooling layer usually follows the convolution layer and performs downsampling to select and filter features, helping to prevent overfitting of the model. Common pooling methods include average pooling and maximum pooling, of which maximum pooling is more common in practical applications. The number is calculated as follows:

$$P_k^j = \max(x_k^j) \qquad (12)$$

In the above formula, $P_k^j$ indicates the final output of layer $j$, $x_k^j$ indicates the layer $j$ area corresponding to maximum pooling.

The activation functions commonly used in deep learning are ReLU functions and Sigmoid functions. The purpose of adding activation functions is to add nonlinear functions to features. ReLU functions can enhance features such as sparsity, and its mathematical expression is as follows:

$$f(a) = \begin{cases} a, & a \geq 0 \\ 0, & a < 0 \end{cases} \qquad (13)$$

The final fully connected layer is used to integrate features and then output features. The output results are classified by Softmax classifier. The mathematical expression is as follows:

$$P(x)_i = \frac{e^{x_i}}{\sum_{i=1}^n e^{x_i}} \qquad (14)$$

In the above formula, $n$ is the classification number, $x_i$ is a feature, $p(x)_i$ is the probability of a feature.

## 2.6 Channel attention

Channel attention[23] (SEnet), originally proposed by Hu et al., can effectively enhance the performance of convolutional neural networks (CNNS) by directly building a model to understand the dependence between channels and dynamically adjusting channel feature feedback. SEnet usually includes squeeze module and excitation module. As shown in the picture below:

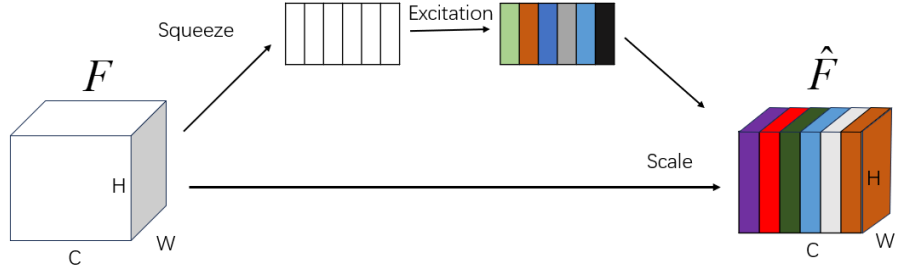

Fig.4 SEnet structure diagram

The Squeeze module compresses two-dimensional features through global averaging pooling to obtain global variables. Global information $z_c$ The mathematical expression is as follows:

$$z_c = F_{sq}(X_c) = \frac{1}{H*W} \sum_{i=1}^{H} X_{c(i,j)} \qquad (15)$$

In the above formula, $X$ is the compressed feature map; $X_c$ is the $c$ channel of the feature graph; $z$ indicates the global information. $z_c$ is the $c$ element of the global information. $H$ and $W$ are the height and width of the input feature map.

The Excitation module assigns a weight value to each channel so that the number of Excitation values in the output is the same as the number of input channels. The weight value $s$ is calculated as follows:

$$s = F_{ex}(z,W) = S(g(z,W)) = S(W_2 R(W_1 z)) \qquad (16)$$

In the above formula, $S$ is the Sigmoid function; $W_1 \in R^{\frac{C}{r} \times C}, W_2 \in R^{C \times \frac{C}{r}}$, $r$ is the reduction rate, and 16 is taken as $r$ in this paper; $R$ is the ReLU function.

In the final feature figure $U_c$, the mathematical formula is as follows:

$$U_c = F_{scale}(u_c, s_c) = u_c \bullet s_c \qquad (17)$$

In the above formula, $s_c$ is the $c$ element of $s$. $u_c \bullet s_c$ represents the product of the channels corresponding to the feature map $u_c$ and the scalar $s_c$.

## 2.7 Model structure

Based on the generative adversarial enhanced multi-scale convolutional neural network model, the main structure consists of three parts, and its structure is shown in Figure 5. Firstly, a small number of one-dimensional time domain signals are converted into two-dimensional images through the Gram difference field. Then, the obtained two-dimensional images are input into the gradient penalty Wasserstein Range-generating Adversarial network (WGAN-GP) model to obtain a large number of data familiar with the original image features. Finally, the obtained data are mixed with the original data and input into the multi-scale attention convolution model. Finally, the fault is identified by softmax classifier.

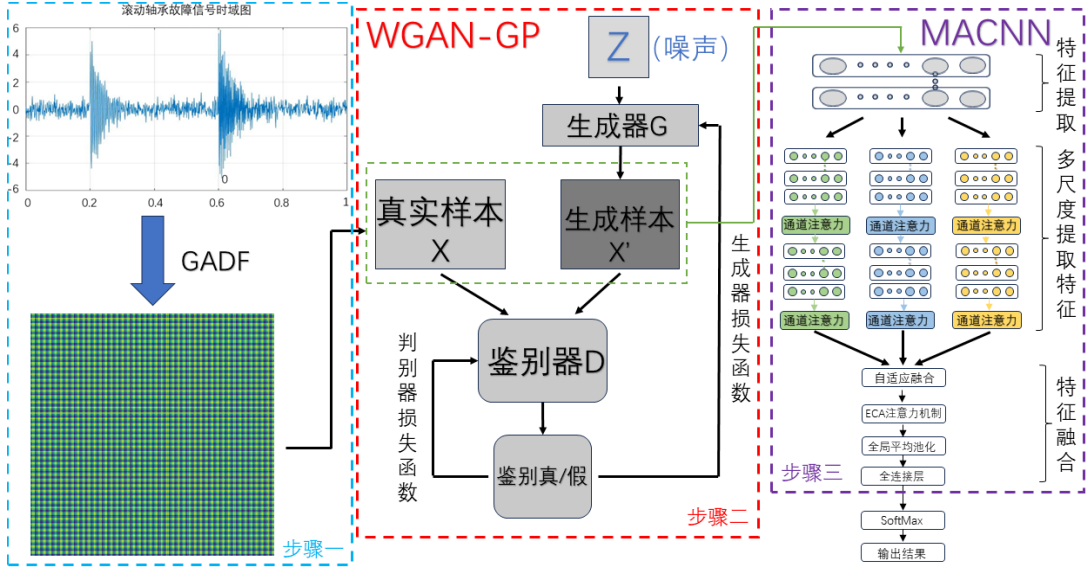

Fig.5 GADF-WGAN-GP-MACNN model

The structure of multi-scale attention convolutional network is shown in Figure 6. Feature maps are first screened by wide convolution layers, which can obtain longer time scale features. Then normalization (IN) is used to reduce the difference between feature graphs, nonlinear activation is realized by ReLU function, and parameter calculation and dimension reduction are reduced by maximum pooling. The mathematical formula is as follows:

$$y_i = P(f(IN(w_i * x_j + b_i))) \qquad (18)$$

In the above formula, $y_i$ is the characteristic output; $w_i$ and $b_i$ are the weight matrix, and $P$ is the maximum pooling.

The multi-scale extraction feature layer is composed of three convolution kernels with different convolution layers and attention mechanisms. Multi-scale convolution can obtain features at different scales, thereby improving the generalization of the model[24]. After feature extraction at different scales, channel weights need to be adjusted by channel attention to further improve the ability of fault feature extraction. Finally, the output dimension is changed by adaptive pooling.

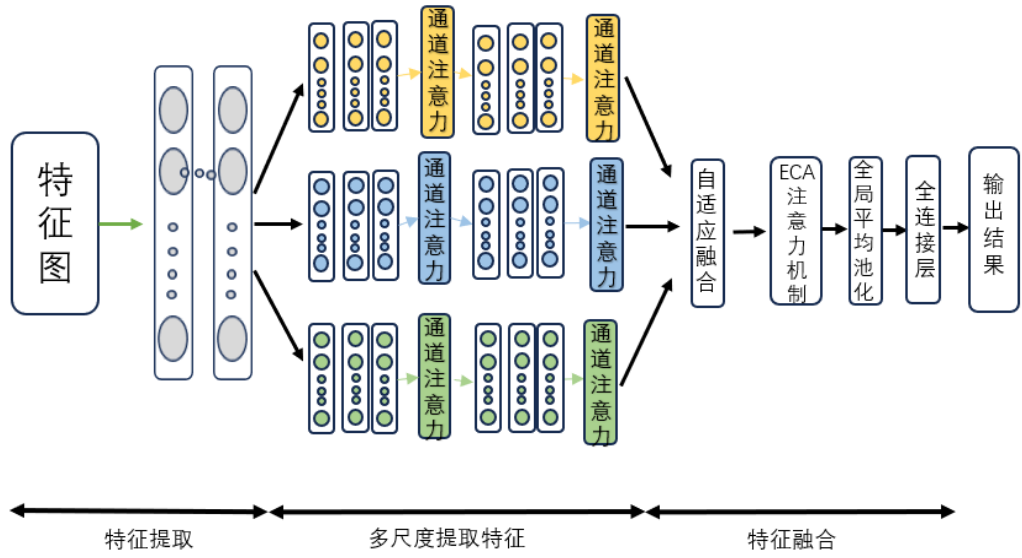

Fig.6 MACNN model

Since the feature fusion of multi-scale convolutional neural networks uses a single channel and a fully connected layer to complete the classification, the feature redundancy is not conducive to the training of the model. Therefore, ECA module is used in the fusion layer for adaptive weighting adjustment, and the adaptive function is as follows:

$$k = \left| \frac{\log_2(C)}{\gamma} + \frac{b}{\gamma} \right| \tag{19}$$

Where, $C$ is the number of channels for input features; $b$ and $\gamma$ are constants.

**Table 1. MACNN Network parameters**

| 层名称 | 核尺寸 | 步长 | 核数 |
| --- | --- | --- | --- |
| 宽卷积层 | 64*1 | 2 | 32 |
| 池化层 | 2*1 | 2 | 32 |
| 多尺度卷积层 | 5*1/7*1/9*1 | 1/1/1 | 64/64/64/ |
| 多尺度池化层 | 2*1 | 2/2/2 | 64/64/64/ |
| 注意力层 | — | — | 64/64/64 |
| 多尺度卷积层 | 5*1/7*1/9*1 | 1/1/1 | 128/128/128 |
| 多尺度池化层 | 2*1 | 2/2/2 | 128/128/128 |
| 注意力层 | — | — | 128/128/128 |
| 全局平均池化 | — | — | — |
| 全连接层 | — | — | — |

## 3 Experimental design and verification

### 3.1 Description of experimental data sets

In order to verify the validity of the model, the XJTU-SY[25] bearing data set is selected as the data set of this experiment. The bearing type is LDK UER204, the sampling frequency is 25.6Hz, 32768 data points are recorded each time, and the sampling period is 1 minute. The data set includes three different working conditions. A total of 15 kinds of bearing start running to failure data, including cage fracture, inner ring failure, outer ring failure, and other failure types, a total of 15 kinds.

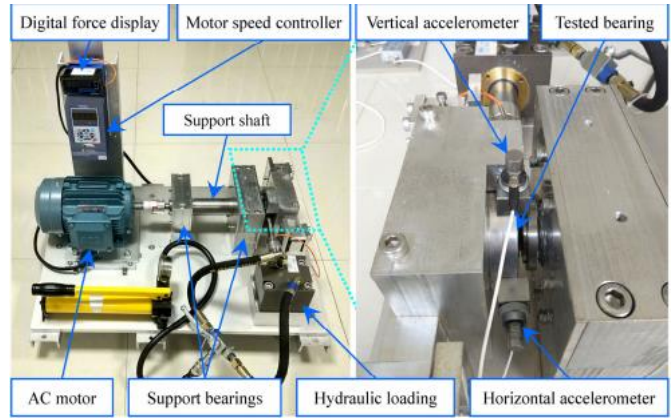

**Fig. 7** Testbed of rolling element bearings

**Table 2.** Parameters of the tested bearings

| Parameter | Value | Parameter | Value |
| --- | --- | --- | --- |
| Outer race diameter | **39.80 mm** | Inner race diameter | **29.30 mm** |
| Load rating (static) | **6.65kN** | Load rating (dynamic) | **12.82kN** |
| Bearing mean diameter | **34.55 mm** | Ball diameter | **7.92 mm** |
| Number of balls | **8** | Contact angle | **0°** |

## 3.2 Experimental verification

2.3.1 Visual training

t-SNE dimensionality reduction technique can be used to visualize the training process of the model and observe the feature distribution before and after training. As shown in the picture below:

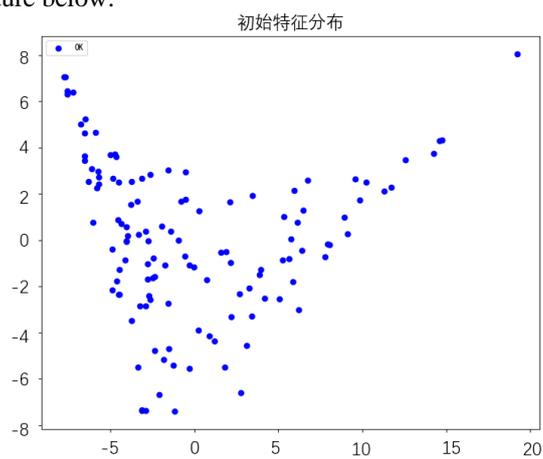

**Fig.8** Initial feature distribution

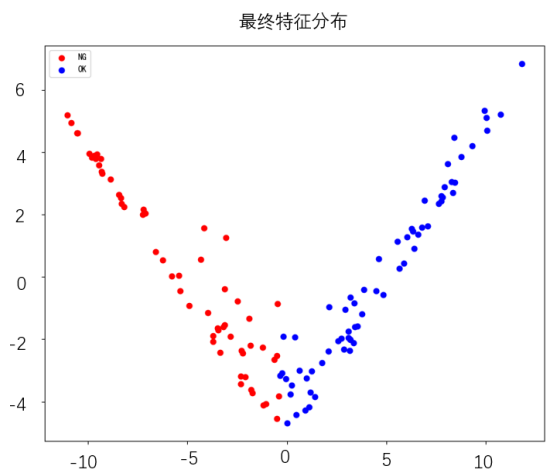

**Fig.9** Final feature distribution

Fig.8 shows the initial distribution, whose feature distribution is cluttered. With the continuous training of the model, the feature distribution in the output distribution Fig.9 is obvious, and the separation is basically realized.

2.3.2 Comparison of different methods

In order to verify the superiority of this model, it is compared with ResNet method. In this model, a convolution layer with a convolution kernel size of 1*5 is selected for the first time, followed by a convolution kernel size of 1*7, and the third convolution kernel size is 1*9. The network depth of ResNet model is 50. Due to the superior ability of ResNet model to extract depth features, it has a strong feature extraction ability. The comparison structure is shown in Figure 11.

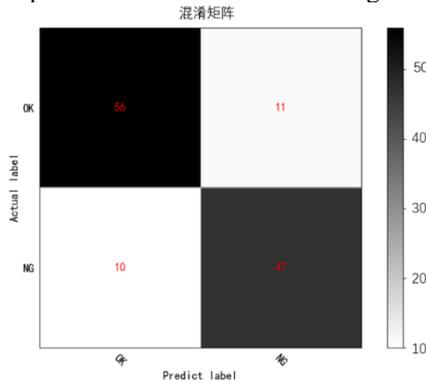 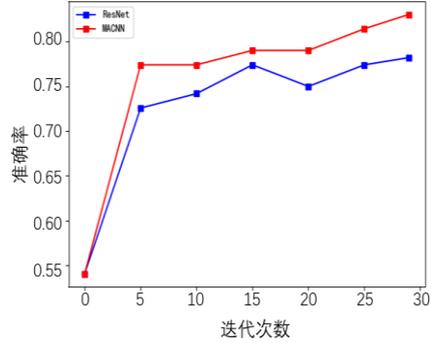

**Fig.10**  Confusion matrix    **Fig.11**  Comparison of different methods

It can be seen that the accuracy of MACNN model is better than that of ResNet model, and the accuracy is about 4% higher. The analysis shows that MACNN has effective fault diagnosis capability.

## 4    Conclusion

In order to solve the problem that current convolutional neural networks can not capture the correlation features between the time domain signals of rolling bearings effectively, and the model accuracy is limited by the number and quality of samples, a rolling bearing fault diagnosis method based on generative adversarial enhanced multi-scale convolutional neural network model is proposed. In this paper, a fault diagnosis method of rolling bearing based on generative adversarial enhanced multi-scale convolutional neural network model is proposed. Firstly, a small number of samples are input into the Gram Angular Difference field (GADF) feature map; Then WGAN-GP network is used to generate a large number of images with similar features to the sample. Finally, the obtained data are input into the multi-scale attention convolutional network model to identify faults. Experiments are carried out on the XJTU-SY bearing data set, and the experiments show that the model has good fault diagnosis ability. Since the experiment is carried out without interference, and there is noise interference in the actual working condition, the bearing fault diagnosis under noise interference and unbalanced data will be studied in the future.

# References

1. Syed, I.; Khadkikar, V.; Zeineldin, H. H., Loss reduction in radial distribution networks using a solid-state transformer. *IEEE Transactions on Industry Applications* **2018,** *54* (5), 5474-5482.
2. Zhao, Z.; Wu, S.; Qiao, B.; Wang, S.; Chen, X., Enhanced sparse period-group lasso for bearing fault diagnosis. *IEEE Transactions on Industrial Electronics* **2018,** *66* (3), 2143-2153.
3. Liu, R.; Yang, B.; Zio, E.; Chen, X., Artificial intelligence for fault diagnosis of rotating machinery: A review. *Mechanical Systems and Signal Processing* **2018,** *108*, 33-47.
4. Xia, M.; Li, T.; Xu, L.; Liu, L.; De Silva, C. W., Fault diagnosis for rotating machinery using multiple sensors and convolutional neural networks. *IEEE/ASME transactions on mechatronics* **2017,** *23* (1), 101-110.
5. Leterme, W.; Azad, S. P.; Van Hertem, D., A local backup protection algorithm for HVDC grids. *IEEE Transactions on Power Delivery* **2016,** *31* (4), 1767-1775.
6. Li, W.; Monti, A.; Ponci, F., Fault detection and classification in medium voltage DC shipboard power systems with wavelets and artificial neural networks. *IEEE Transactions on Instrumentation and Measurement* **2014,** *63* (11), 2651-2665.
7. Chen, Z.; Mauricio, A.; Li, W.; Gryllias, K., A deep learning method for bearing fault diagnosis based on cyclic spectral coherence and convolutional neural networks. *Mechanical Systems and Signal Processing* **2020,** *140*, 106683.
8. Liu, S.; Yang, N.; Li, M.; Zhou, M. In *A recursive recurrent neural network for statistical machine translation*, Proceedings of the 52nd Annual Meeting of the Association for Computational Linguistics (Volume 1: Long Papers), 2014; pp 1491-1500.
9. Zhao, H.; Jiang, L.; Jia, J.; Torr, P. H.; Koltun, V. In *Point transformer*, Proceedings of the IEEE/CVF international conference on computer vision, 2021; pp 16259-16268.
10. Goodfellow, I.; Pouget-Abadie, J.; Mirza, M.; Xu, B.; Warde-Farley, D.; Ozair, S.; Courville, A.; Bengio, Y., Generative adversarial nets. *Advances in neural information processing systems* **2014,** *27*.
11. Wang, X.; He, H.; Li, L., A hierarchical deep domain adaptation approach for fault diagnosis of power plant thermal system. *IEEE Transactions on Industrial Informatics* **2019,** *15* (9), 5139-5148.
12. Ding, Y.; Ma, L.; Ma, J.; Wang, C.; Lu, C., A generative adversarial network-based intelligent fault diagnosis method for rotating machinery under small sample size conditions. *IEEE Access* **2019,** *7*, 149736-149749.
13. 肖雄; 肖宇雄; 张勇军; 宋国明; 张飞, 基于二维灰度图的数据增强方法在电机轴承故障诊断的应用研究. *中国电机工程学报* **2021,** *41* (2), 738-749.
14. Jiang, G.; He, H.; Yan, J.; Xie, P., Multiscale convolutional neural networks for fault diagnosis of wind turbine gearbox. *IEEE Transactions on Industrial Electronics* **2018,** *66* (4), 3196-3207.
15. 张明德; 卢建华; 马婧华, 基于多尺度卷积策略 CNN 的滚动轴承故障诊断. *重庆理工大学学报(自然科学)* **2020,** *34* (6), 102-110.
16. 彭鹏; 柯梁亮; 汪久根, 噪声干扰下的 RV 减速器故障诊断. *机械工程学报* **2020,** *56* (1), 30-36.
17. 沈长青; 王旭; 王冬; 阙红波; 石娟娟; 朱忠奎, 基于多尺度卷积类内迁移学习的列车轴承故障诊断. *交通运输工程学报* **2020,** *20* (5), 151-164.

18.Xiao, F.; Chen, Y.; Zhu, Y., GADF/GASF-HOG: feature extraction methods for hand movement classification from surface electromyography. *Journal of Neural Engineering* **2020,** *17* (4), 046016.

19.骆家杭；张旭；汪靖翔, 基于格拉姆角场和多尺度 CNN 的轴承故障诊断. *轴承* **2022,** *6*, 73-78.

20.Adler, J.; Lunz, S., Banach wasserstein gan. *Advances in neural information processing systems* **2018,** *31*.

21.邵海东；李伟；刘翔；杨斌, 时变转速下基于双阈值注意力生成对抗网络和小样本的转子-轴承系统故障诊断. *机械工程学报* **2023,** *59* (12), 215-224.

22.Ulyanov, D.; Vedaldi, A.; Lempitsky, V. In *Improved texture networks: Maximizing quality and diversity in feed-forward stylization and texture synthesis*, Proceedings of the IEEE conference on computer vision and pattern recognition, 2017; pp 6924-6932.

23.Hu, J.; Shen, L.; Sun, G. In *Squeeze-and-excitation networks*, Proceedings of the IEEE conference on computer vision and pattern recognition, 2018; pp 7132-7141.

24.An, Z.; Li, S.; Wang, J.; Xin, Y.; Xu, K., Generalization of deep neural network for bearing fault diagnosis under different working conditions using multiple kernel method. *Neurocomputing* **2019,** *352*, 42-53.

25.Wang, B.; Lei, Y.; Li, N., Xjtu-Sy bearing datasets. Dec: 2018.